\documentclass[fleqn,twoside]{article}
\usepackage{espcrc2}

\usepackage{amssymb}
\usepackage{amsfonts}
\usepackage{amsmath}

\usepackage[all]{xy}
\usepackage{graphicx}
\usepackage[figuresright]{rotating}

\title{Simple current extensions and permutation orbifolds in string theory.}

\author{M. Maio\address[Nikhef]{Nikhef, Science Park 105,\\ 
        1098 XG Amsterdam, The Netherlands}%
        \thanks{Extended version of the Proceedings of the Cargese Summer School 2010 
         ``Formal Developments and Applications'', 
         Nucl. Phys. B (Proc. Suppl.) {\bf 216} (2011) 252.}}

\begin{document}

\begin{abstract}
We review extensions by integer spin simple currents in two-dimensional conformal field theories
and their applications in string theory. In particular, we study the problem of resolving the
fixed points of a simple current and apply the formalism to the permutation orbifold.
In terms of string compactifications, we construct permutations of $N=2$ minimal models
and use them as building blocks in heterotic Gepner models.\\

\indent NIKHEF/2011-023
\vspace{1pc}
\end{abstract}

\maketitle

\section{Simple currents and fixed points}

(Rational) Conformal field theories \cite{Belavin:1984vu} not only play a special role within String Theory, but they are also interesting objects in their own right, since they appear in many other contests such as in condensed matter physics. \\
Standard ways are known to derive new conformal field theories from existing ones. The prototype is the \textit{orbifold} CFT of the form $G/H$: given a current algebra $G$, one mods out its symmetry subalgebra $H$, leaving only $H$-invariant states plus twisted fields, necessary for modular invariance. Much is known about these coset models.\\
Another prototype example is the extension via \textit{integer spin simple currents} (see \cite{Schellekens:1990xy} for a review). Extensions are very powerful tools in String Theory, since they allow to perform projections (e.g. GSO projection), impose constraints (such as the $\beta$-constraints in Gepner models) or implement field identifications in coset models. An extension is an orbifold-like procedure, that is possible to apply when the CFT has got \textit{simple currents} with integer spin. \\
A simple current $J$ is by definition a special field of the theory with simple fusion rules,
\begin{equation}
(J)\times(i)\equiv(Ji)\,,
\nonumber
\end{equation}
having only one contribution on the r.h.s. In practice, what one does in extensions is modding out the discrete symmetry generated by $e^{2\pi i Q_J}$, being $Q_J(i)$ the monodromy charge of the field $i$ with respect to the simple current $J$. One is left with zero-charge \textit{orbits} under $J$ of the form $(i,Ji,J^2i,\dots,J^{N-1}i)$, for some integer $N$. $N$ is called the order of $J$. It can happen sometimes that $Jf\equiv f$ for some field $f$. $f$ will be then called a \textit{fixed point} of J.\\
Fixed points are very delicate objects to handle. In the new, or \textit{extended}, theory each fixed point gives rise to \textit{splitted} fields, in number equal to the order of the current, on which there is a priori no control. As a consequence, the extended $S$ matrix, $\tilde{S}$, cannot be immediately expressed in terms of the $S$ matrix of the original theory. Instead, one has to introduce a set of new matrices, the so-called $S^J$ matrices, one for each simple current $J$, in terms of which $\tilde{S}$ is parametrized as \cite{Fuchs:1996dd}:
\begin{equation}
\tilde{S}_{(a,i)(b,j)}=\frac{|G|}{\sqrt{|U_a||S_a||U_b||S_b|}}\sum_{J\in G}\Psi_i(J) S^J_{ab} \Psi_j(J)^{\star}\,.
\nonumber
\end{equation}
In this formula, the index $(a,i)$ labels the $i^{th}$ field into which $a$ is splitted; the prefactor is a group-theoretical quantity, acting as a normalization; the $\Psi_i(J)$ are the group characters acting as phases; the sum is over all the simple currents used to extend the original theory. \\
The $S^J$ matrices act only on fixed points of $J$, i.e. $S^J_{ab}=0$ if either $a$ or $b$ is not fixed by $J$. Moreover, modular invariance of the full $\tilde{S}$ matrix implies modular invariance of $S^J$:
\begin{equation}
S^J\cdot (S^J)^\dagger=1\,, \qquad (S^J\cdot T^J)^3=(S^J)^2\,.
\nonumber
\end{equation}
Hence, in this formalism, the problem of determining the extended $S$ matrix is equivalent to finding the set of $S^J$ matrices. This is known as the \textit{fixed point resolution} problem.

\section{The permutation orbifold}
Theories for which the $S^J$ matrices were already known in the past include WZW models and coset theories. Recently, the fixed point problem has been solved for the cyclic permutation orbifold, in the simplified case of two factors:
\begin{equation}
 \mathcal{A}_{\rm perm} \equiv \mathcal{A}\times \mathcal{A}/\mathbb{Z}_2\,.
\nonumber
\end{equation}
the $\mathbb{Z}_2$ exchanging the two factors. $\mathcal{A}$ denotes an arbitrary CFT.\\
The field content of the generic permutation orbifold was already known from \cite{Klemm:1990df}. It consists of three kinds of fields: \textit{diagonal}, $\phi_{(i,\chi)}$, \textit{off-diagonal}, $\phi_{\langle m,n\rangle}$ with $m<n$, and \textit{twisted}, $\phi_{\widehat{(i,\chi)}}$. Also the $S$ matrix of this orbifold was already known from \cite{Borisov:1997nc} in terms of the $S$ and $T$ matrices of the original CFT $\mathcal{A}$. We will denote it by $S^{BHS}$.\\
If one is interested in performing extensions of this cyclic orbifold, the first thing that must be asked is whether or not it admits simple currents and, in affirmative case, if the simple currents have fixed points. It turns out that the answers to both questions is yes \cite{Maio:2009kb,Maio:2009cy,Maio:2009tg}. Orbifold simple currents are diagonal fields corresponding to the simple currents of the original CFT, while their fixed points can be of any kind. Hence, one has to resolve those fixed points or, equivalently, determine the corresponding $S^J$ matrices. One possible strategy is to give an \textit{ansatz} for $S^J$ which satisfies modular invariance and furthermore is subject to the following constraints:
\begin{itemize}
\item for the identity current, $J=0$, $S^J$ must reduce to $S^{BHS}$;
\item the extension by the anti-symmetric component of the identity undoes
      the permutation orbifold, giving back the tensor product 
      $\mathcal{A}\times \mathcal{A}$ \cite{Maio:2009kb};
\item the $S^J$ must be consistent with known expressions existing for some
      WZW models with particular current algebras (e.g. $A(1)$, $B(n)$, $D(2n)$)
      \cite{Maio:2009kb,Maio:2009cy}.
\end{itemize}
The explicit expression for the ansatz is given in \cite{Maio:2009tg}. There, it is also shown that unitarity and modular invariance are satisfied. Moreover, integrality of the fusion rules has been checked numerically for very large rational CFT's, even though its proof is probably doable. All these non-trivial checks strongly suggest that this is indeed the correct answer.

\section{$N=2$ minimal models and string compactifications}
The previously derived formalism has been applied to four dimensional string theory compactifications in the context of Gepner models \cite{Gepner:1987qi,Gepner:1987vz}. Gepner models are built out of tensor products of an external four-dimensional spacetime part and an internal $c=9$ conformal field theory with $N=2$ worldsheet supersymmetry. The internal sector itself consists of tensor products of $r$ $N=2$ minimal models, where additional constraints are imposed in order to guarantee worldsheet and spacetime supersymmetry. These constraints are just extensions of the tensor product by specific integer-spin simple currents. In the $i^{\rm th}$ minimal model ($i=1,\dots,r$) the relevant fields are the worldsheet supersymmetry current $T_F$ with weight $h=\frac{3}{2}$ and the spectral flow operator $S_F$ with weight $h=\frac{c_i}{24}$. The currents that impose the Gepner constraints are then
\begin{equation}
W_i=(0,\dots,0,T_F,0,\dots,0;V) \qquad i=1,\dots,r
\nonumber
\end{equation}
for the worldsheet supersymmetry, where $T_F$ appears in position $i$, and 
\begin{equation}
S_{\rm susy}=(S_F,\dots,S_F,\dots,S_F;S)
\nonumber
\end{equation}
for the spacetime supersymmetry. Here, $V$ and $S$ denote the vector and spinor currents of the $SO(10)_1$ which is introduced by the bosonic string map.

If two\footnote{For more than two identical factors, the $S^J$ matrices are not currently available.} of the factors in the tensor product are identical, we can then replace them by their permutation orbifold. Performing extensions and permutations in the product of two identical minimal models, we discover that \cite{Maio:2010eu}:
\begin{itemize}
\item the current $T_F\otimes T_F$ makes the tensor product supersymmetric on the worldsheet;
\item the current $(T_F,1)$ is the worldsheet supersymmetry current of the permutation orbifold, making the permutation orbifold supersymmetric on the worldsheet; the following scheme summarizes the structure:
\begin{displaymath}
\xymatrix{
\boxed{(N=2)^2} \ar@/^/[d]^{BHS} \ar[r]^{T_F\otimes T_F} & \boxed{(N=2)^2_{\rm Susy}} \ar@/^/@2{->}[d]^{\rm super-BHS} \\
\boxed{(N=2)^2_{\rm orb}} \ar@/^/@{.>}[u]^{(0,1)} \ar[r]^{(T_F,1)\quad} & \boxed{(N=2)^2_{\rm Susy-orb}} \ar@/^/@{.>}[u]^{(0,1)}
}
\end{displaymath}
here ``super-BHS'' denotes the analogous BHS mechanism carried out in the supersymmetric tensor product.
\item the current $(T_F,0)$ is \textit{not} the worldsheet supersymmetry current of the permutation orbifold, but extends it into a new non-supersymmetric CFT $X$; the summarizing graph is below:
\begin{displaymath}
\xymatrix{
\boxed{(N=2)^2} \ar@/^/[d]^{BHS} \ar[r]^{T_F\otimes T_F} & \boxed{(N=2)^2_{\rm Susy}}  \\
\boxed{(N=2)^2_{\rm orb}}  \ar@/^/@{.>}[u]^{(0,1)} \ar[r]^{(T_F,0)\quad} & \boxed{{\rm Non-Susy}\,\,X} \ar[u]^{(0,1)}
}
\end{displaymath}
\item in the $(T_F,\psi)$-extended orbifold ($\psi=0,1$), ``exceptional'' simple currents appear as a consequence of the extension procedure; they are exceptional in the sense that they generate from off-diagonal fields in the mother CFT, hence a priori not expected; moreover, in some cases, they admit fixed points, which must then be resolved\footnote{However their resolution is currently an open problem.}.
\end{itemize}

When considering string compactifications, we can now use the permutation orbifold of minimal models (and their extensions) to construct new four dimensional theories \cite{Maio:2011qn}. The oldest approach is to use the heterotic string, where families of chiral fermions come out automatically in representations of $SO(10)$-based GUT models. By breaking the GUT gauge symmetry group one can study various features, such as the presence of fractionally-charged particles and the abundance of three-family models. The relevant results are the following.
\begin{itemize}
\item The gauge groups that are considered are the GUT group $SO(10)$ and seven rank-5 subgroups, namely:
\begin{itemize}
\item the Pati-Salam group $SU(4)\times SU(2)\times SU(2)$,
\item the Georgi-Glashow GUT group $SU(5)\times U(1)$,
\item two global realizations of the left-right symmetric algebra $SU(3)\times SU(2)\times SU(2)\times U(1)$, called $(LR, Q=1/3)$ and $(LR, Q=1/6)$,
\item three global realizations of the standard model algebra $SU(3)\times SU(2)\times U(1)\times U(1)$, called $(SM, Q=1/2)$, $(SM, Q=1/3)$ and $(SM, Q=1/6)$.
\end{itemize}
The last two Lie algebras are distinguished by their fractional charged representations (hence $Q$ denotes the quantization of the elementary charge). These eight gauge groups are obtained as simple current extensions of the standard model affine Lie algebra $SU(3)_{1}\times SU(2)_1\times U(1)_{30}$.
\item Despite the fact that no evidence for fractional charged particles exists in nature, with a limit of less then $10^{-20}$ per nucleon, generically they are known to appear, and they do appear, in string theory when the GUT group is broken down to subgrups by modding out freely acting discrete symmetries. These particles can only be vector-like and must have a huge mass (Planck or string scale). Fractional charged particles with masses of the order of the TeV scale would be a serious problem, unless they are charged under some additional non-abelian gauge group which confines them.
\item Models with three families are normally disfavored compared to two or four. The generic trend is that 
\begin{itemize}
\item even number of families is always more favorable than an odd,
\item these distributions decrease exponentially when the number of families increases.
\end{itemize}
Figure \ref{famplot_standard} shows the family distribution for the permutation orbifolds of standard Gepner models. All the models (no exceptional invariants are considered) have even number of families. In particular, the number three is missing. In order to improve the abundance of three-family models, one can consider suitable ``lifts''. The lifting procedure is described in \cite{GatoRivera:2009yt} and consists in replacing one of the minimal models, together with the $E_8$ factor, by a different CFT with the same modular invariant properties. This procedure in general raises the weights of the primary fields and produces many spectra with odd number of generations.
\begin{figure}[p]
\begin{center}
\includegraphics[scale=0.3]{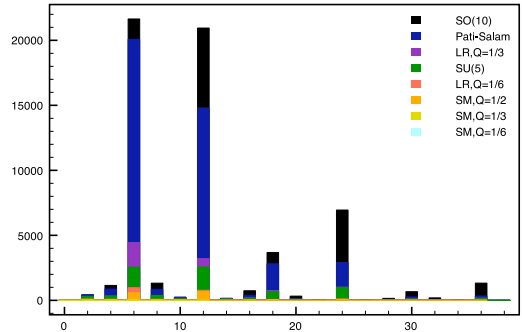}
\caption{\small Distribution of the number of families for permutation orbifolds of standard Gepner Models.}
\label{famplot_standard}
\end{center}
\end{figure}
Figure \ref{famplot_lift} shows the family distribution for lifted Gepner models.
\begin{figure}[p]
\begin{center}
\includegraphics[scale=0.3]{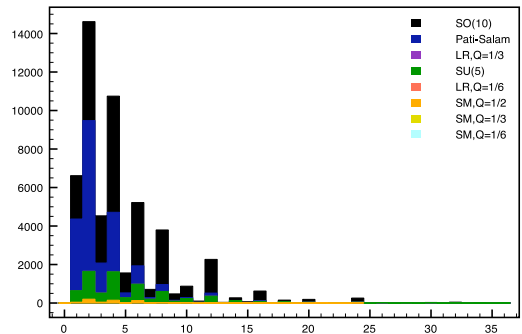}
\caption{\small Distribution of the number of families for permutation orbifolds of lifted Gepner Models.}
\label{famplot_lift}
\end{center}
\end{figure}
\noindent \\
A similar lifting procedure can be carried on with the additional $U(1)$ which accompanies the standard model gauge group and is often related to a B-L symmetry \cite{GatoRivera:2010fi}. In this case, two lifts are possible (lift A and lift B). Figures \ref{famplot_liftA} and \ref{famplot_liftB} refer to such cases.
\begin{figure}[p]
\begin{center}
\includegraphics[scale=0.3]{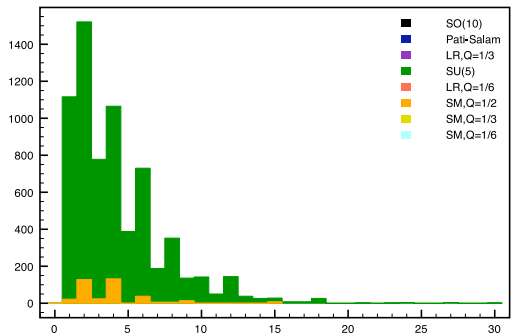}
\caption{\small Distribution of the number of families for permutation orbifolds of B-L lifted (lift A) Gepner Models.}
\label{famplot_liftA}
\end{center}
\end{figure}
\begin{figure}[p]
\begin{center}
\includegraphics[scale=0.3]{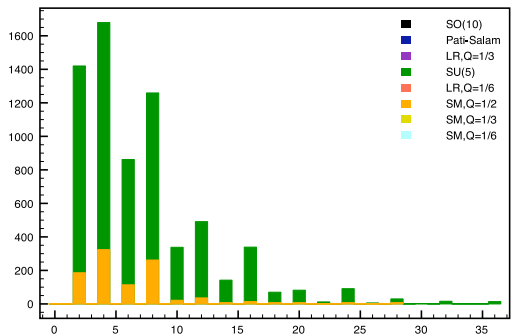}
\caption{\small Distribution of the number of families for permutation orbifolds of B-L lifted (lift B) Gepner Models.}
\label{famplot_liftB}
\end{center}
\end{figure}
\end{itemize}

\section*{Acknowledgments}
Research supported by the Dutch Foundation for Fundamental Research of Matter (FOM) as part of the program STQG (String Theory and Quantum Gravity). MM would like to thank the organizers of the Cargese Summer School 2010 for creating a very stimulating environment.

\end{document}